\documentclass[10pt,conference]{IEEEtran}
\usepackage{color}
\usepackage{amsmath}
\usepackage{amssymb}
\usepackage{graphicx}

\IEEEoverridecommandlockouts   

\renewcommand{\v}{\mathbf}
\newcommand{\wv}[1]{{\widehat{\v{#1}}}}

\newcommand{\V}{V}
\newcommand{\M}{M}

\def\slantfrac#1#2{\tiny {\hbox{$\,^#1\!\!/\!_#2$}}}
\newcommand{\oh}{{\slantfrac{1}{2}}}
\renewcommand{\c}{{\mathsf{c}}}

\begin{document}


\title{
The E8 Lattice and Error Correction in \\ Multi-Level Flash Memory
}

\author{\IEEEauthorblockN{Brian M.~Kurkoski} 
\IEEEauthorblockA{University of Electro-Communications\\ 
Tokyo, Japan\\ 
kurkoski@ice.uec.ac.jp} 
\thanks{This research was supported in part by the Ministry of Education, Science, Sports and Culture; Grant-in-Aid for Scientific Research (C) number 21560388.}%
}


\maketitle

\begin{abstract}
A construction using the E8 lattice and Reed-Solomon codes for error-correction in flash memory is given.   Since E8 lattice decoding errors are bursty, a Reed-Solomon code over GF($2^8$) is well suited.   This is a type of coded modulation, where the Euclidean distance of the lattice, which is an eight-dimensional signal constellation, is combined with the Hamming distance of the code.  This system is compared with the conventional technique for flash memories, BCH codes using Gray-coded PAM.   The described construction has a performance advantage of 1.6 to 1.8 dB at a probability of word error of $10^{-6}$.  Evaluation is at high data rates of 2.9 bits/cell for flash memory cells that have an uncoded data density of 3 bits/cell.
\end{abstract}

\section{Introduction}

While single-level flash memory stores a single bit per memory cell, data density can be increased by using two, three or more bits per cell \cite{Nobukata-ssc00} \cite{Grossi-ed03}.    Along with this increase in density, the influence of noise also increases.   Numerous error-correcting approaches have been considered, although BCH codes are predominant in practice.  

A conventional multilevel-flash chip uses Gray-coded PAM and presents hard decisions externally.  Because the signal-to-noise ratio in flash can be characterized as high, Gray-coded PAM results in single bit errors.   Further, errors in flash memories tend to be uncorrelated.   Thus, the errors may be characterized as random, isolated errors of one bit, and flash memory systems (such as SSD) use BCH error correcting codes of high rate.  The decoder in such chips is a hard-input BCH decoder, implemented in an external chip.


In carrier-based AWGN systems, two-dimensional constellations such as QPSK and QAM are used.   Trellis-coded modulation is a low-complexity method to improve the performance by combining the Euclidean distance of the constellation with the Hamming distance of a convolutional code, or a more powerful error-correcting code \cite{Ungerboeck-it82}. In fact, trellis-coded modulation with convolutional codes and a QAM constellation (PAM over two dimensions) has been considered for general memories \cite{Lou-cds00}, and have been evaluated for flash memories \cite{Sun-CSD07}. In carrier-based systems, two-dimensional constellations aid with synchronization at the receiver, but this is not needed with flash memories.  

This paper considers the use of lattices --- that is, higher dimensional constellations --- for error-correction in flash memories.   In particular, the E8 lattice, which is the best-known lattice in eight dimensions, is considered for flash memory.   While E8 has greater minimum Euclidean distance than PAM, an outer error-correcting code is still needed to guarantee data reliability.   Whereas trellis-coded modulation uses a convolutional code, in this paper Reed-Solomon codes, which are a type of block code, are used.

Because E8 decoding induces burst-like errors, Reed-Solomon (RS) codes constructed over GF($2^8$) are used for error correction.   One RS code symbol is encoded to eight memory cells; a group of eight memory cells represents one lattice point.  When an E8 decoding error occurs, with high probability it will be one of the 240 neighboring lattice points.  Thus, the constellation (lattice) to codeword mapping needs only to distinguish between these neighbors.   Only the modulo-2 value of the lattice points are protected by the RS codes; the Euclidean separation of the lattice points is also important.   

The E8 lattice has a number of desirable properties.   It is the the best-known lattice in eight dimensions, in the sense of having the densest packing, highest kissing number and being the best quantizer \cite[p.~12]{Conway-1999}.  It also has an efficient decoding algorithm.  The lattice generator matrix is triangular,  which makes it suitable for encoding.    In addition, the E8 lattice points are either integers or half-integers; for implementations, this may be more suitable than writing arbitrary values to memory.

Previously, the E8 lattice has been considered for trellis-coded modulation with a convolutional code and average power shaping \cite{Calderbank-procieee86}.  In addition, there are numerous proposals for including RS codes in trellis-coded modulation \cite{Wei-com94} \cite{Laneman-broad01}.   
However, the combination of the E8 lattice and RS codes described in this paper appears to be unique.  For error-correction in flash memories, BCH codes have received the most attention; however recently RS codes, which can be constructed over a smaller field for the same block length, have also been considered for flash memory \cite{Chen-08}.  At high rates, the rate loss of RS codes compared to BCH codes is relatively small, and the smaller field size makes RS decoding more efficient.

\section{Background}

This section gives some background.  First the assumed channel model is described, followed by a brief overview of lattices in general.   Then, the E8 lattice and some of its properties are described.

\subsection{Channel Model}

The assumed model has $N$ flash memory cells, which can store an arbitrary value between 0 and $\V$.   The uncoded information rate is $\log_2 q$ bits per cell.   In conventional multilevel flash memory, this is accomplished by choosing $\V = q-1$ and storing one of $\{0,1,\ldots,q-1\}$ in each cell.   For lattices of dimension $n$, the lattice encodes $q^n$ levels, or $n \log_2 q$ bits in $n$ cells.

The flash reading and writing process introduces noise, which is modeled as additive white Gaussian noise with mean zero and variance $\sigma^2$. A more advanced model, which assigns higher variances to the levels $0$ and $q-1$ is difficult to apply to lattice coding, since the cells store continuous values (see for example \cite{Sun-CSD07}).  The channel SNR used in this paper is:
\begin{eqnarray}
\frac{\V^2}{\sigma^2},
\end{eqnarray}
where $\V^2$ represents the peak signal energy.

\subsection{Lattices and Lattice Codes}

An $n$-dimensional lattice $\Lambda$ is an infinite set of points $\mathbf x = (x_1,x_2, \ldots, x_n)^t$ defined by an $n$-by-$n$ generator matrix $G$, for which
\begin{eqnarray}
\mathbf x &=& G \mathbf b, \label{eqn:lattice}
\end{eqnarray}
where $\mathbf b =(b_1,\ldots, b_n)^t$ is from the set of all possible integer vectors, $b_i \in \mathbb Z$.  The $i,j$ entry of $G$ is denoted $g_{ij}$.    The set $\Lambda$ forms a group under addition in $\mathbb R^n$, so lattices are linear in the sense that the sum of any two lattice points is a lattice point.   


The minimum of the squared Euclidean distance between any two distinct lattice points is the minimum norm.   The lattice points at this distance from the origin are the minimum vectors.  The number of points at this distance is called the kissing number, denoted $\tau$.   Because of the linearity of lattices, all lattice points have the same number of nearest neighbors, corresponding to the minimum vectors.    The packing radius $\rho$ is taken to be half the square root of the minimum norm.   In terms of finite-field codes, the minimum vectors are analogous to the minimum-weight non-zero codes words, and the packing radius is analogous to half the minimum distance. When the lattice is scaled by $\alpha$, its generator matrix is $\alpha G$ and the packing radius is $\alpha \rho$.

A practical coding scheme must select a finite subset of the lattice;  a codebook may be constructed by the intersection of a shaping region $B$ and the lattice $\Lambda$.  In general, codebook construction is difficult, but when $B$ is an $n$-cube and $G$ is a triangular matrix, there exists a practical method \cite{Sommer-itw09} \cite{Kurkoski-globesub}.  In particular, let $B$ be an $n$-cube, given by:
\begin{eqnarray}
& 0 \leq x_i < \M,\label{eqn:encoding}
\end{eqnarray}
and let $\v a = (a_1,\ldots, a_n)$ be information-containing integers, with
\begin{eqnarray}
a_i &\in& \{0,1,\ldots, \frac{\M}{g_{ii}}-1 \}
\end{eqnarray}
for $i=1,\ldots,n$.   The diagonal elements must satisfy the condition that $M/g_{ii}$ is an integer.  It is convenient to assume that $\M / g_{ii}$ is a power of two, to aid encoding from bits to integers.

In general the lattice point $G \cdot \v a$ is not in $B$.   Instead, the encoding finds a vector $\v k = (\frac{k_1}{g_{11}}, \ldots, \frac{k_n}{g_{nn}})$, which determines $\v b$:
\begin{eqnarray}
\mathbf b &=& \mathbf a + M \mathbf k, 
\end{eqnarray}
such that 
\begin{eqnarray}
\mathbf x = G \cdot \mathbf b
\end{eqnarray}
is in the cube $B$.
Because the generator matrix is lower triangular, the $k_i$ can be found by solving the inequality (\ref{eqn:encoding}):
\begin{eqnarray}
0 \leq & \sum_{j=0}^{i-1} g_{ji} b_j + g_{ii} \big( a_i + \frac{M}{g_{ii}} k_i ) & < \M 
\end{eqnarray}
for $k_i$, which is unique.   From the triangular structure of $G$, first $k_1$, then $k_2, \ldots, k_n$ are found in sequence.   In particular:
\begin{eqnarray}
k_i &=& \Big\lceil \frac{
 - \sum_{j=0}^{i-1} g_{ji} b_j - g_{ii} a_i
}{
M 
}
\Big\rceil,
\end{eqnarray}
where the computation at step $i$ depends upon $b_1,\ldots, b_{i-1}$.   

Since the power constraint for flash memory is cubic, one might expect that this encoding scheme is sufficient, by choosing $\M=\V$.   Unfortunately, the strict upper inequality in (\ref{eqn:encoding}) is necessary for $k_i$ to be unique.  As a result, it is not possible to find lattice points for which $x_i = \M$, which reduces the size of the codebook.   But this problem is resolved for the E8 lattice in the next section.

Decoding is straightforward.   If $\widehat{\v x}$ is a lattice point, and $\widehat{\v b} = G^{-1} \widehat{\v x}$, then $\widehat{\v a}$ is found as:
\begin{eqnarray}
\widehat a_i &=& \widehat b_i \mod \frac{\M}{g_{ii}}.
\end{eqnarray}

\subsection{E8 Lattice}

\def\ahalf{\small {\hbox{$\mathbf{\frac 1 2}$}}}

The E8 lattice is an eight-dimensional lattice, which can be described in terms of the D8 checkerboard lattice \cite[p.~120]{Conway-1999}.   The D8 lattice points are all integers that have an even sum.   The E8 lattice is the union of the D8 lattice, and a coset of the D8 lattice:
\begin{eqnarray}
\textrm{E8} &=& \textrm{D8}  \cup \textrm{D8} + {\textstyle \mathbf{\frac 1 2}}
\end{eqnarray}
where,
\begin{eqnarray}
{\textstyle\mathbf{\frac 1 2}} &=& ( \oh, \oh, \oh, \oh, \oh, \oh, \oh, \oh ).
\end{eqnarray}

A generator matrix is:
\begin{eqnarray}
G &=&
\left[
\begin{array}{rrrrrrrr}
\oh & 0 & 0 & 0 & 0 & 0 & 0 & 0 \\
\oh & 1 & 0 & 0 & 0 & 0 & 0 & 0 \\
\oh & -1 & 1 & 0 & 0 & 0 & 0 & 0 \\
\oh & 0 & -1 & 1 & 0 & 0 & 0 & 0 \\
\oh & 0 & 0 & -1 & 1 & 0 & 0 & 0 \\
\oh & 0 & 0 & 0 & -1 & 1 & 0 & 0 \\
\oh & 0 & 0 & 0 & 0 & -1 & 1 & 0 \\
\oh & 0 & 0 & 0 & 0 & 0 & -1 & \phantom{-}2 \\
\end{array}
\right] . \label{eqn:egen}
\end{eqnarray}

The kissing number $\tau$ of the E8 lattice is 240.    The minimal vectors are sequences $(\pm 1^2, 0^6)$ (there are $4\cdot {8 \choose 2}$ such sequences) and $(\pm \oh,\pm \oh,\pm \oh,\pm \oh,\pm \oh,\pm \oh,\pm \oh,\pm \oh)$, where there are an even number of minus signs (there are $2^7$ such sequences).   The packing radius $\rho$ is $1 / \sqrt{2}$.

The E8 lattice has an efficient decoding algorithm, which finds the lattice point $\wv x$ closest to an arbitrary point $\v y \in \mathbb R^8$ \cite{Conway-it82}.   The decoding algorithm is low-complexity: it requires only rounding operations, wrong-way rounding of the least reliable position, and comparison of Euclidean distance.

\section{Proposed Construction}

This section describes the proposed construction.   First,  the codebook is  created by scaling the lattice by $\alpha < 1$. Then, the encoding is given, which partitions the flash memory into blocks of 8 cells, each encoded using an E8 lattice.  Each block corresponds to one RS code symbol.   Finally, the decoding algorithm is described, including a post-processing step which uses the lattice error patterns to recover the integer sequence.

\subsection{Lattice Codebook}

The lattice code is the set of lattice points which satisfy the constraint that the cell value is between 0 and $\V$ (inclusive), and cell values correspond to lattice points.  A significant point is that the encoding described previously does not allow encoding $x_i = \M$, but the physical system allows $x_i = \V$.  The lattice should be scaled so that the lattice code (that is, $B \cap \Lambda$)  contains $q^8$ lattice points.  For the E8 lattice, it is observed that choosing
\begin{eqnarray}
\alpha &=& \frac{\V}{\V+0.5} \textrm{ and} \\
\M &=& \V+1
\end{eqnarray}
will satisfy this rate criteria and the power constraint.   To show this, consider an example with $q=4$ and $\V=3$.   Initially encode the unscaled E8 lattice using not $\M=3$, but $\M=4$; this violates the power constraint.   Because the E8 lattice consists of integer points and half-integer points, this encoding results in lattice points with a maximum value of 3.5. By scaling the lattice by $3 / 3.5$ the half-integer lattice points are placed exactly on the boundary $\V$.   

Using the unscaled lattice, the total number of lattice codewords that satisfy $0 \leq x_i \leq \V+1$ is for the E8 lattice is:
\begin{eqnarray}
\frac{(\V+1)^n}{ | \det G|} &=& (\V+1)^n = q^n.
\end{eqnarray}
The scaling does not change the number of lattice codewords.  For convenience, encoding is done with the unscaled lattice, (\ref{eqn:egen}), and scaling is applied before writing to memory.   

The scaling reduces the separation between lattice points.   For the E8 lattice and large $q$, $\rho$ approaches the maximum value of $1/\sqrt{2}$.   For the small value of $q=4$, $\alpha = 6/7$ results in an effective packing radius of 0.606; this is higher 0.5, the packing radius of the PAM constellation.

\subsection{Encoding}

\newcommand{\X}{{\mathsf x}}
\renewcommand{\a}{{\mathsf a}}
\newcommand{\nc}{{n_{\mathrm c}}}
\newcommand{\kc}{{k_{\mathrm c}}}

Consider $N$ flash memory cells.   These $N$ cells are separated into blocks, each block consists of 8 cells, so there are $N/8$ blocks.   Let $\X$ denote the vector of blocks, that is:
\begin{eqnarray}
\X &=& (\v x_1, \v x_2, \ldots, \v x_{N/8}),
\end{eqnarray}
where each $\v x_i$ is a lattice point.

A $(\nc, \kc, t)$ shortened, systematic RS code constructed over $\textrm{GF}(2^8)$ is used.   Each RS code symbol is assigned to a block, so $\nc = N/8$.   Blocks to which systematic RS symbols and parity symbols are assigned are called systematic blocks and parity blocks, respectively.  The block $\v x_i$ is systematic for $i = 1,\ldots,\kc$ and is parity for $i=\kc,\ldots,\nc$.   The  information integers are represented as:
\begin{eqnarray}
\a &=& \big(\v a_1, \v a_2, \ldots, \v a_{\kc}, \v a_{\kc+1}, \ldots, \v a_{\nc} \big),
\end{eqnarray}
written to separate the integers in systemic blocks and integers in parity blocks.

Encoding for systematic blocks is identical to the general lattice encoding described previously.  For an example of $q=8$, within any systematic block, if $\v a$ is the vector of information integers, then
 $a_1 \in \{0,1,\ldots, 15\}$,  $a_8 \in \{0,1,2,3\}$ and the  remaining six integers are from $\{0,1,\ldots,7\}$.

Encoding for parity blocks requires combining the computed RS parity checks and additional information integers. For each systematic block, compute:
\begin{eqnarray}
\v u_i = \v a_i \mod 2,
\end{eqnarray}
for $i=1,\ldots,\kc$, where $\v u = \v a \mod 2$ means component-wise modulo-2.
The eight bits $(u_1,\ldots,u_8)$ form a single GF($2^8$) RS symbol.   Using information $\v u_1,\ldots,\v u_{k \c}$, compute the RS parity symbols $\v p_1, \ldots, \v p_{n \c - k \c}$.    For each parity block, $\v p = (p_1,p_2,\ldots,p_8), p_i \in \{0,1\}$.

To perform lattice encoding in the parity blocks, integers are formed where the least-significant bit is the RS parity, and the remaining part is information.   The information integers in the parity block are $a_i \in \{0,1,\ldots, \frac{M}{2 g_{ii}} \}$.   For the example of $q=8$: $a_1 \in \{0,1,\ldots, 7\}$,  $a_8 \in \{0,1\}$ and the  remaining six integers are from $\{0,1,2,3\}$. The integers to be lattice encoded in parity block $i$ are:
\begin{eqnarray}
\v p_{i-\kc} + 2 \cdot \v a_{i},
\end{eqnarray}
for $i=\kc+1,\ldots,\nc$, and addition is over the real numbers. Thus, additional information is embedded in the parity blocks, because only the LSB of the parity block is needed by the RS parity symbol.   Note also that Gray coding is not used.

The total number of encoded information bits is $k$,
\begin{eqnarray}
k &=&  \kc  \cdot 8 \log_2 q + (\nc - \kc) \big( -8 + 8\log_2 q )
\end{eqnarray}
and the total number of cells used is $N =  8 n_{\c}$, so the code rate $R=k/N$, measured in bits per cell is:
\begin{eqnarray}
R &=&  \frac{\kc}{\nc} \log_2 q + \frac{ \nc - \kc}{\nc} \big( -1 + \log_2 q ) \textrm{ bits/cell}. \label{eqn:rate}
\end{eqnarray}

\newcommand{\x}{r}
\begin{table*}
\begin{centering}
\setlength{\tabcolsep}{4pt}%
\begin{tabular}{\x\x\x\x\x\x\x\x | rrrrrrrr | cccccccc}
\multicolumn{8}{c|}{$\wv x -\v x$} &
\multicolumn{8}{c|}{$\wv a- \v a$} &
\multicolumn{8}{c}{$\wv u \oplus \v u$}  \\ \hline \hline \\
\multicolumn{24}{c}{error patterns of type $(\pm 1^2, 0^6)$ } \\ \hline
 -1 &  -1 &   0 &   0 &   0 &   0 &   0 &   0 &  -2 &   0 &   1 &   2 &   3 &   4 &   5 &   3 &   0 &   0 &   1 &   0 &   1 &   0 &   1 &   1 \\
  1 &   1 &   0 &   0 &   0 &   0 &   0 &   0 &   2 &   0 &  -1 &  -2 &  -3 &  -4 &  -5 &  -3 &   0 &   0 &   1 &   0 &   1 &   0 &   1 &   1 \\
 -1 &   1 &   0 &   0 &   0 &   0 &   0 &   0 &  -2 &   2 &   3 &   4 &   5 &   6 &   7 &   4 &   0 &   0 &   1 &   0 &   1 &   0 &   1 &   0 \\
  1 &  -1 &   0 &   0 &   0 &   0 &   0 &   0 &   2 &  -2 &  -3 &  -4 &  -5 &  -6 &  -7 &  -4 &   0 &   0 &   1 &   0 &   1 &   0 &   1 &   0 \\
  1 &  -1 &   0 &   0 &   0 &   0 &   0 &   0 &   2 &  -2 &  -3 &  -4 &  -5 &  -6 &  -7 &  -4 &   0 &   0 &   1 &   0 &   1 &   0 &   1 &   0 \\
 -1 &   1 &   0 &   0 &   0 &   0 &   0 &   0 &  -2 &   2 &   3 &   4 &   5 &   6 &   7 &   4 &   0 &   0 &   1 &   0 &   1 &   0 &   1 &   0 \\
  1 &   1 &   0 &   0 &   0 &   0 &   0 &   0 &   2 &   0 &  -1 &  -2 &  -3 &  -4 &  -5 &  -3 &   0 &   0 &   1 &   0 &   1 &   0 &   1 &   1 \\
 -1 &  -1 &   0 &   0 &   0 &   0 &   0 &   0 &  -2 &   0 &   1 &   2 &   3 &   4 &   5 &   3 &   0 &   0 &   1 &   0 &   1 &   0 &   1 &   1 \\
 -1 &   0 &  -1 &   0 &   0 &   0 &   0 &   0 &  -2 &   1 &   1 &   2 &   3 &   4 &   5 &   3 &   0 &   1 &   1 &   0 &   1 &   0 &   1 &   1 \\
  1 &   0 &   1 &   0 &   0 &   0 &   0 &   0 &   2 &  -1 &  -1 &  -2 &  -3 &  -4 &  -5 &  -3 &   0 &   1 &   1 &   0 &   1 &   0 &   1 &   1 \\
 -1 &   0 &   1 &   0 &   0 &   0 &   0 &   0 &  -2 &   1 &   3 &   4 &   5 &   6 &   7 &   4 &   0 &   1 &   1 &   0 &   1 &   0 &   1 &   0 \\
  1 &   0 &  -1 &   0 &   0 &   0 &   0 &   0 &   2 &  -1 &  -3 &  -4 &  -5 &  -6 &  -7 &  -4 &   0 &   1 &   1 &   0 &   1 &   0 &   1 &   0 \\
\multicolumn{8}{c}{\vdots} &
\multicolumn{8}{c}{\vdots} &
\multicolumn{8}{c}{\vdots} 
 \\
\\
\multicolumn{24}{c}{error patterns of type $(\pm \oh, \cdots \pm \oh)$ } \\ \hline
-$\oh$ & -$\oh$ & -$\oh$ & -$\oh$ & -$\oh$ & -$\oh$ & -$\oh$ & -$\oh$ &  -1 &   0 &   0 &   0 &   0 &   0 &   0 &   0 &   1 &   0 &   0 &   0 &   0 &   0 &   0 &   0 \\
$\oh$ & $\oh$ & $\oh$ & $\oh$ & $\oh$ & $\oh$ & $\oh$ & $\oh$ &   1 &   0 &   0 &   0 &   0 &   0 &   0 &   0 &   1 &   0 &   0 &   0 &   0 &   0 &   0 &   0 \\
$\oh$ & -$\oh$ & -$\oh$ & -$\oh$ & -$\oh$ & -$\oh$ & -$\oh$ & $\oh$ &   1 &  -1 &  -2 &  -3 &  -4 &  -5 &  -6 &  -3 &   1 &   1 &   0 &   1 &   0 &   1 &   0 &   1 \\
-$\oh$ & $\oh$ & $\oh$ & $\oh$ & $\oh$ & $\oh$ & $\oh$ & -$\oh$ &  -1 &   1 &   2 &   3 &   4 &   5 &   6 &   3 &   1 &   1 &   0 &   1 &   0 &   1 &   0 &   1 \\
-$\oh$ & $\oh$ & -$\oh$ & -$\oh$ & -$\oh$ & -$\oh$ & -$\oh$ & $\oh$ &  -1 &   1 &   1 &   1 &   1 &   1 &   1 &   1 &   1 &   1 &   1 &   1 &   1 &   1 &   1 &   1 \\
$\oh$ & -$\oh$ & $\oh$ & $\oh$ & $\oh$ & $\oh$ & $\oh$ & -$\oh$ &   1 &  -1 &  -1 &  -1 &  -1 &  -1 &  -1 &  -1 &   1 &   1 &   1 &   1 &   1 &   1 &   1 &   1 \\
$\oh$ & $\oh$ & -$\oh$ & -$\oh$ & -$\oh$ & -$\oh$ & -$\oh$ & -$\oh$ &   1 &   0 &  -1 &  -2 &  -3 &  -4 &  -5 &  -3 &   1 &   0 &   1 &   0 &   1 &   0 &   1 &   1 \\
-$\oh$ & -$\oh$ & $\oh$ & $\oh$ & $\oh$ & $\oh$ & $\oh$ & $\oh$ &  -1 &   0 &   1 &   2 &   3 &   4 &   5 &   3 &   1 &   0 &   1 &   0 &   1 &   0 &   1 &   1 \\
-$\oh$ & -$\oh$ & $\oh$ & -$\oh$ & -$\oh$ & -$\oh$ & -$\oh$ & $\oh$ &  -1 &   0 &   1 &   1 &   1 &   1 &   1 &   1 &   1 &   0 &   1 &   1 &   1 &   1 &   1 &   1 \\
$\oh$ & $\oh$ & -$\oh$ & $\oh$ & $\oh$ & $\oh$ & $\oh$ & -$\oh$ &   1 &   0 &  -1 &  -1 &  -1 &  -1 &  -1 &  -1 &   1 &   0 &   1 &   1 &   1 &   1 &   1 &   1 \\
$\oh$ & -$\oh$ & $\oh$ & -$\oh$ & -$\oh$ & -$\oh$ & -$\oh$ & -$\oh$ &   1 &  -1 &  -1 &  -2 &  -3 &  -4 &  -5 &  -3 &   1 &   1 &   1 &   0 &   1 &   0 &   1 &   1 \\
-$\oh$ & $\oh$ & -$\oh$ & $\oh$ & $\oh$ & $\oh$ & $\oh$ & $\oh$ &  -1 &   1 &   1 &   2 &   3 &   4 &   5 &   3 &   1 &   1 &   1 &   0 &   1 &   0 &   1 &   1 \\
-$\oh$ & $\oh$ & $\oh$ & -$\oh$ & -$\oh$ & -$\oh$ & -$\oh$ & -$\oh$ &  -1 &   1 &   2 &   2 &   2 &   2 &   2 &   1 &   1 &   1 &   0 &   0 &   0 &   0 &   0 &   1 \\
$\oh$ & -$\oh$ & -$\oh$ & $\oh$ & $\oh$ & $\oh$ & $\oh$ & $\oh$ &   1 &  -1 &  -2 &  -2 &  -2 &  -2 &  -2 &  -1 &   1 &   1 &   0 &   0 &   0 &   0 &   0 &   1 \\
\multicolumn{8}{c}{\vdots} &
\multicolumn{8}{c}{\vdots} &
\multicolumn{8}{c}{\vdots} 
 \\
\end{tabular}
\caption{Several of the 240 minimum norm error patterns for the E8 lattice.  The vector $\widehat{\v u} \oplus \v u$ is sufficient to identify the error pattern, except for a sign change.\label{table:errorpatterns}}
\end{centering}
\end{table*}

\subsection{Decoding}

The encoded lattice point is scaled by $\alpha = \V / (\V + 0.5)$, passed through an AWGN channel, and scaled by $1/\alpha$ by the decoder.   Let this received sequence be denoted by $\big(\v y_1, \v y_2, \ldots, \v y_{\nc}\big)$.   Block-by-block E8 lattice decoding is performed  \cite{Conway-it82}, and the integer sequence $\big( \wv{a}_1, \ldots, \wv{a}_\nc\big)$ is obtained.
Compute $\wv{u} = \wv{a} \mod 2$ and perform RS decoding.    Consider any single block.  If lattice decoding is successful, then the RS symbol will be correct.  However, if lattice decoding is unsuccessful, with high probability, a minimal vector error is made, that is, the lattice decoder erroneously selects one of the 240 neighboring lattice points.   

If the number of block errors (that is, lattice decoding errors) is less than the error-correcting capability $t$ of the RS code, then the blocks which have errors can be identified.  But since the RS code only protects the least-significant bits, RS decoding alone cannot recover all the information.    The distance properties of the lattice will be used.   This may be regarded as a type of post-processing step.

The difference $\wv a - \v a$ denotes the integer error pattern, and $\wv u \oplus \v u$ = $| \wv u - \v u| $ denotes the bit error pattern.   Since $\v u$ is provided by the RS decoder, $\wv u \oplus \v u$ is known. A sample of some of the 240 error patterns is given in Table~\ref{table:errorpatterns}.  Except for a sign change throughout, the bit error pattern $\wv u \oplus \v u$ uniquely identifies the integer error pattern.   By employing a look-up table (much like Table~\ref{table:errorpatterns}) this bit error pattern $\wv u \oplus \v u$ can be mapped to an integer error pattern $\v e$, used to find a two new estimated integer sequences $\wv a_1$ and $\wv a_2$:
\begin{eqnarray}
\wv{a}_1 &=& \wv{a} +\v e, \textrm{ and}\\
\wv{a}_2 &=& \wv{a} -\v e.
\end{eqnarray}
Select the vector $\widehat{\wv a}$ that has the shortest distance from the received sequence $\v y$:
\begin{eqnarray}
\widehat{\wv a} &=& \left\{\begin{array}{ll}
\wv a_1 & ||\wv a_1 - \v y|| \leq || \wv a_2 - \v y|| \\ 
\wv a_2 & \textrm{otherwise}
\end{array} \right. .
\end{eqnarray}
Thus, the entire integer sequence can generally be recovered, if the correct RS symbol is known.

\begin{table*}
\begin{center}
\begin{tabular}{cccccc|cccccccc}
\multicolumn{6}{c|}{RS over GF($2^{8}$) } & \multicolumn{6}{c}{BCH over GF($2^{13}$) } \\[4pt] \hline
                       &         &                   &               &  RS rate &  Flash rate, bits/cell &                 &         &                   &                                       & BCH Rate & Flash rate, bits/cell\\
$(\nc,\kc,t)$  & $s$ & cells $N$ & bits $k$ & $\kc / \nc$  & $R$  (\ref{eqn:rate}) & $(\nc,\kc,t)$  &  $s$ & cells $N$ & bits $k$ & $\kc / \nc$ & $\kc \log_2 q / \nc$ \\[1pt] \hline \hline
 (172,170,1) &83 & 1376 & 4112  &  0.988  & 2.988   & (4109,4096,1) &4082 & 1370 & 4096 & 0.997 & 2.991 \\
(172,168,2) &83 & 1376 & 4096  &   0.977  &   2.977 & (4122,4096,2) &4069 & 1374 & 4096  & 0.994 & 2.981 \\
(173,167,3) &82 & 1384 & 4104  &   0.965  &   2.965 & (4135,4096,3) &4056 & 1379 & 4096  & 0.991 & 2.972 \\
(174,166,4) &81 & 1392 & 4112  &   0.954  &   2.954 & (4148,4096,4) &4043 & 1383 & 4096  & 0.987 & 2.962\\
(174,164,5) &81 & 1392 & 4096  &    0.943 &   2.943 & (4161,4096,5) &4030 & 1387 & 4096  & 0.984 & 2.953
\end{tabular}
\end{center}
\caption{RS and BCH codes considered in Fig.~\ref{fig:fig1} \label{table:codes}}
\end{table*}

\section{Numerical Results}

The proposed construction is evaluated numerically.  Comparisons are made with the dominant conventional system for error-correction in flash memories, BCH codes using Gray-coded PAM. To show the benefit of the proposed construction, a fair comparison is made by selecting code parameters to encode about 4096 bits, so that the number of cells $N$ (and thus the code rate) are as similar as possible.

For parameters $m$, $t$ and $s$, there exists a systematic, shorted RS code of length $\nc = 2^{m} -1-s$ symbols, encoding $\kc = 2^m - 1 - 2t-s$ information symbols, that can correct $t$ symbol errors.   
For parameters $m$, $t$ and $s$, there exists a systematic, shortened BCH code of length $ 2^{m} -1-s$ bits, encoding $ 2^m - 1 - mt-s$ information bits, that can correct $t$ symbol errors.  

\begin{figure}
\begin{center}
\includegraphics[width=9cm]{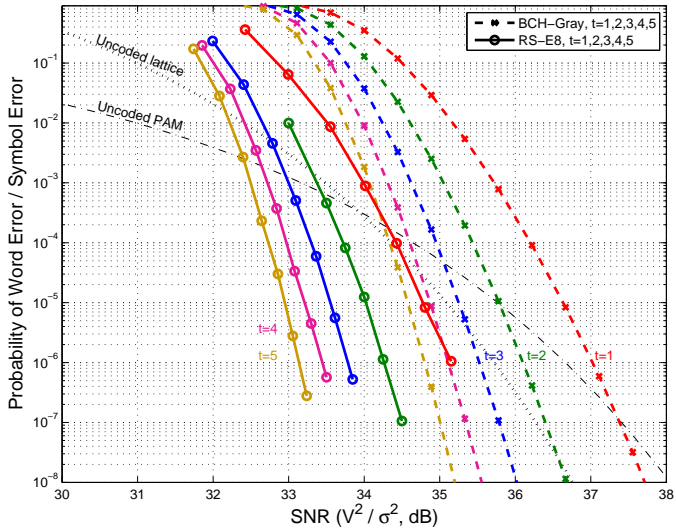}
\end{center}
\caption{Probability of word error for Reed-Solomon codes with E8 lattice, compared to BCH codes with Gray-coded PAM, for flash memory with $q=8$ (uncoded 3 bits/cell).   The proposed construction has 1.6 to 1.8 dB better performance. The uncoded performance is also shown, where an error corresponds to a symbol (an E8 lattice symbol or a q-ary PAM symbol) being in error.}
\label{fig:fig1}
\end{figure}

Fig.~\ref{fig:fig1} shows the simulated word error probability for $q=8$ (3 bits per flash cell), and various code rates.   The specific code parameters are in Table \ref{table:codes}.   At a probability of word error rate of $10^{-6}$, the uncoded E8 lattice has approximately 1.8 dB better performance than PAM.   This benefit is preserved after coding, with gains of 1.6 to 1.8 dB are observed.  Note that for each code comparison the error-correction capability of the RS and BCH code is essentially the same.  

It should be noted that the E8 lattice decoder is a soft-input decoder, whereas PAM decoding is hard decision.  In both cases, the RS and BCH decoders are using hard inputs.   In commercial flash memory products, the memory and error-correction functions in separate chips.   While the E8 decoder has soft inputs, because it is a low-complexity decoder, future flash memory systems could implement a lattice decoder on the flash memory chip.   Thus, the error-correction performance may be improved, without significant changes to existing system architecture. 




\end{document}